\documentclass[doublecol]{epl2} 

\usepackage{amsmath}
\usepackage{amssymb}
\usepackage{bm}
\usepackage{nicefrac}
\usepackage[square,sort,comma,numbers]{natbib}
\usepackage{cleveref}
\usepackage{graphicx}
\usepackage{subcaption}

\renewcommand{\vec}[1]{\bm{{#1}}}
\renewcommand{\d}{\;\mathrm{d}}

\newcommand{\pderiv}[3][]{\frac{\partial^{#1}#2}{\partial #3^{#1}}}
\newcommand{\deriv}[3][]{\frac{\mathrm{d}^{#1}#2}{\mathrm{d}#3^{#1}}}
\DeclareMathOperator{\re}{Re}
\DeclareMathOperator{\im}{Im}

\newcommand{\pR}{\phi^R}
\newcommand{\pL}{\phi^L}
\newcommand{\pJ}{\phi^J}

\newcommand{\mathcl}[1]{#1}

\usepackage{xcolor}

\title{Stay in your lane: Density fluctuations in multi-lane traffic}

\author{J. Worsfold\inst{1} \and T. Rogers\inst{1}}
\shortauthor{J. Worsfold \etal}

\institute{                    
  \inst{1} Centre for Mathematical Biology, Department of Mathematical Sciences, The University of Bath, Claverton Down, Bath BA2 7AY\\
}

\abstract{
When a new vehicle joins a lane, those behind may have to temporarily slow to accommodate them. Changing lane can be forced due to lane drops or junctions, but may also take place spontaneously at discretion of drivers, and recent studies have found that traffic jams and traffic oscillations can form even without such bottlenecks.
Understanding how lane changing behaviour affects traffic flow is important for learning how to design roads and control traffic more effectively.
Here, we present a stochastic model of spontaneous lane changing which exhibits a reduction in the overall flow of traffic. 
By examining the average flow rate both analytically and through simulations we find a definitive slow down of vehicles due to random switching between lanes. By extending the model to three lane traffic we find a larger impact on the flow of the middle lane compared to the side lanes.
}

\begin{document}

\maketitle

\section{Introduction}

It is generally understood that when a vehicle changes lanes on a multi-lane road, it impacts the surrounding vehicles and can reduce the flow of traffic \cite{originalMultilane,KernerExperimental}.
Recent work \cite{ZhengOscillations} has found that oscillations and substantial reductions in traffic flow can often be attributed to lane changing manoeuvres rather than decisions based on following those in front.
Understanding the mechanisms and effects of lane changing is thus a key ingredient to attempts to reduce congestion.
Originally, work on multi-lane flow centred around continuum models of traffic. 
The earliest models \cite{originalMultilane,MichalopoulosMacroModel} can be captured by a generic PDE for the flow of the density,
\begin{align}
    \partial_t \rho_i + \partial_x q_i = s_i 
    \label{eq:ogmodel}
\end{align}
where density and flow along lane $i$ are given by $\rho_i$, $q_i$, respectively. The term $s_i$ on the right hand side represents the net flux into the lane $i$ and can be a function of both the density and flow rate but was originally just a linear function of the density. 
In essence, equations of this type encode first-order models like the paradigmatic Lighthill-Whitham model \cite{Whitham-Lighthill,helbing2001traffic}, modified by the addition of flux between lanes.

Without a reason to change lanes in homogeneous flowing traffic, $s_i$ would be zero and so no lane changes would occur. 
Real-world experience of multi-lane traffic contradicts this as drivers regularly change lanes in relatively free-flowing traffic. 
This consistent low level of unpredictable activity can be approximated by random lane changes modelled as Poisson processes. 
In \cite{LavalHybrid2006}, this was used in tests for lane dropping scenarios and found to be negligible. 
Other recent work has also focused on systematic lane-changing areas such as preparing for junctions or lane drops \cite{OhBottlenecks}. 
For instance, Jin \cite{JinBottlenecks} considered a position-dependent lane changing intensity.
The time taken to change lanes coupled with this intensity creates a higher effective density since space is taken up in both lanes as the manoeuvre occurs. 
Away from bottlenecks where systematic lane changes are not present, however, the effect of unpredictable lane changes has not been quantified. Where such behaviour has been considered in the literature, the focus has been on complex simulations \cite{NextGenSimulation} or microscopic interactions \cite{MicroProbLaneChange}, not on extracting theoretical predictions of the impact on traffic flow.

Typically, microscopic models either build on Totally Asymmetric Simple Exclusionary Processes (TASEP) \cite{ASEP,helbing2001traffic}, often referred to as cellular automata models \cite{CAReview,NagelCA}, or study the gaps between cars called car following models \cite{CFReview,GippsCF}.
By choosing a different formulation for the movement and interaction of vehicles, we are able to circumvent many of the difficulties encountered when incorporating lane changes into the aforementioned models.

We start from an individual, particle based depiction of traffic with vehicles moving continuously along the road according to an interaction kernel with nearby vehicles but with random changes between lanes. 
From this we find a mathematical description of the density in each lane akin to that mentioned in \eqref{eq:ogmodel}. 
Crucially, however, the stochastic effects from the individuals and finite nature of the system are retained.
Our main result is that the impact of this noise in the density profile is a measurable decrease in the average velocity of each vehicle, on average.
While the initial result focuses on the case of two lanes, we go further to show how the flow is reduced in three lane traffic.
In this case the higher flux in and out of the middle lane induces a greater effect on the velocity when compared to the side lanes.

\section{Stochastic model}

To begin to understand the influence of unpredictable decisions on the overall flow of traffic, we start from a minimal representation of unpredictable lane changing and vehicle interactions. The movement of the vehicles will be captured in two forms: continuous forward motion and discrete changes of lane. There are a fixed total number of vehicles, $N$, although the number of vehicles in each lane is not fixed. The vehicles are indexed $j=1,\dots,N$ and the sets $\mathcl{L,R,M}$ contain the indices of the vehicles in the left, right and middle lanes respectively at any moment in time. Since we are interested in behaviour away from bottlenecks, we impose periodic boundary conditions to ensure inflow and outflow do not affect the calculations. We chose a (non-dimensional) domain of length $2\pi$ for computational simplicity, and specify vehicle locations by $x_j\in[0,2\pi)$. While moving within a lane, vehicles have (again non-dimensional) maximum velocity $1$, but will reduce their velocity if needed to maintain separation from the traffic ahead. The amount they slow is determined by the distribution of vehicles and a coupling kernel $\sum_{j\in J}K(x_i-x_j)/N$ for a vehicle, $i$, in lane $J$. The kernel is chosen such that individuals do not consider vehicles behind them ($K(x)=0\,,x>0$). The analysis we present can be followed for any sensible choice of kernel, though we typically make a particular choice of an exponential kernel in order to solve some integrals exactly. Specifically, we use
    \begin{align}
        K(x) = \frac{\beta}{\alpha}\exp\left(\frac{x}{\alpha }\right), \quad x<0\;.
    \end{align}
    and scale $\alpha=4\pi m/N$ so the total slow down is independent of the population size and $m\in\mathbb{N}$ indicates the number of cars within a typical lengthscale effecting the drivers velocity. We model changes of lane as instantaneous events occurring random times according to independent Poisson processes of rate $\lambda$. While the processes are independent, the rate is common to all vehicles. 

We shall show that this simple model can be considered as a non-local version of the Lighthill-Whitham model if we are ambivalent to the lane dynamics and if the system is large. For intermediate system sizes, it will emerge that the lane switching plays a key factor in the overall flow of the traffic. 

\begin{figure}
    \centering
    \includegraphics[width=8cm]{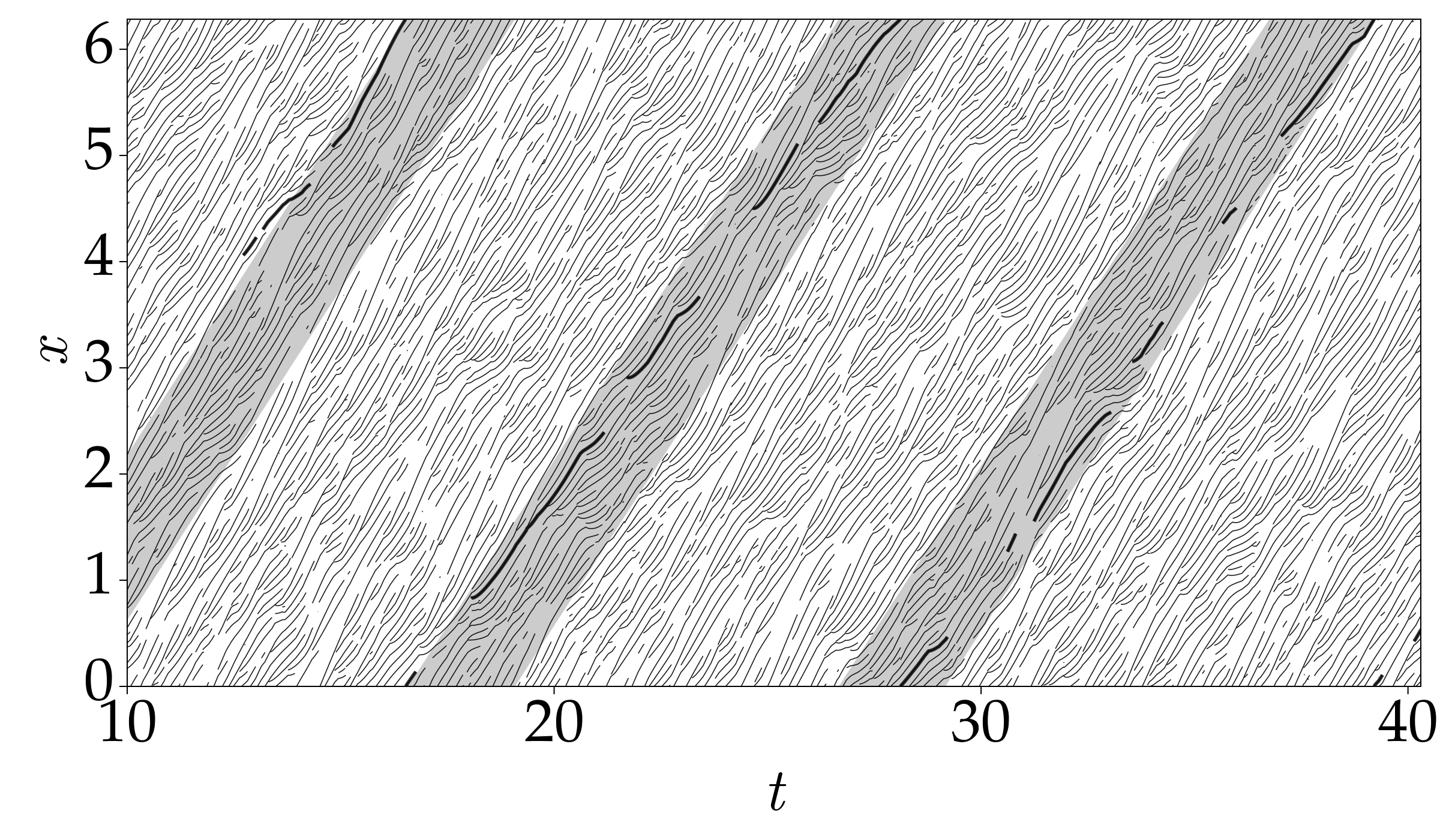}
    \caption{Trace plot of $N=100$ vehicles in the left lane moving slower on average than the velocity $V^*$ achieved when lane switching is prevented, indicated with grey stripes. A typical vehicle trace is highlighted in black, missing sections correspond to periods spent in the other lane. }
    \label{fig:traceplot}
\end{figure}

As can be seen from \Cref{fig:traceplot}, there is inherent variability to the velocities of the vehicles as they regularly adjust to the increased or decreased density of vehicles in front of them. When a lane change occurs, there is often a cascade of vehicles behind which reduce their velocity as they approach this slightly higher density region. Since the interaction kernel is continuous and long-ranged, eventually the vehicles adjust to the new spacing or another lane change occurs.

\section{Expansion of the Master equation}

In order to move from an individual description of the vehicles to an overall picture of the flow rate, we introduce the lane-dependent densities:
\begin{align}
    \phi^J(x) = \frac{1}{N}\sum_{j\in \mathcl{J}}\delta(x-x_j)
\end{align}
where, for now, 
$J\in\{L,R\}$ refers to either the left or right lanes. The model is first order with respect to the movement of vehicles, in particular they move at a rate dependent on the others in their lane. The definition of lane densities allows us to write this succinctly as 
\begin{align*}
    \dot{x}_i & = 1-\frac{1}{N}\sum_{j\in \mathcl{J}} K(x_i-x_j),\quad i\in \mathcl{J} \\
    & = 1 - (K\star\phi^J)(x_i)
\end{align*} 
where we define the convolution as $(K\star\phi)(x)=\int_{0}^{2\pi}\phi(y)K(x-y)\d y$. We proceed by writing a full master equation for the density states. Because the positional information is retained when a vehicle switching lanes, it is useful to define the step operators
\begin{align}
    \Delta_y^{L\pm}F(\pL,\pR) = F\left(\pL(x)\pm\frac{1}{N}\delta(x-y),\pR\right)
\end{align}
and similarly for $\Delta_y^{R\pm}$. There are two components to this system: stochastic switching of lane; and deterministic transport of vehicles along the road. The Master equation is broken down into these two components so they can be treated separately. Mathematically, the probability density of a density state is written as $P(\vec{\phi})$ and evolves according to the rate of lane switching $Q_\lambda(\vec{\phi},x)$ and vehicle transport $Q_{h}(\vec{\phi},x,y)$ such that
\begin{align*}
    \partial_t P(\vec{\phi}) = N\iint \left[Q_{\lambda}(\vec{\phi},x) + Q_{h}(\vec{\phi},x,y)\right]P(\vec{\phi}) \d x \d y\;.
\end{align*}
For lane switching, individuals change species type but retain their position and so is local to the position $x$. In terms of the step operators, this can be summarised as 
\begin{align*}
    Q_{\lambda}(\vec{\phi},x) = &\left(\Delta_x^{L-}\Delta_x^{R+}-1\right)\lambda\pL(x) \\ &+ \left(\Delta_x^{R-}\Delta_x^{L+}-1\right)\lambda\pR(x)
\end{align*}
Incorporating a deterministic flow into the Master equation is done by approximating continuous movement of individuals with discrete jumps. These jumps are of size $hu(\phi^J,x)$ where $u(\phi^J,x)$ is the instantaneous velocity at a point $x$ while the timestep, $h$, is the reciprocal of the rate at which these jumps occur. The resulting term in the Master equation is 
\begin{align}
    Q^J_{h}(\vec{\phi},x,y) = (\Delta_x^{J-}\Delta_y^{J+}-1)d(x-y)\phi^J(x)
\end{align}
where $d(x-y)=\frac{1}{h}\delta(x+u(\phi^J,x)h-y)$ is the jump rate density.

Now, we employ a Kramers-Moyal expansion \cite{van1992stochastic} in orders of $1/N$ on the step operators such that they are given by their Taylor series
\begin{align}
    \Delta_x^{J\pm} = 1 \pm \frac{1}{N}\frac{\delta}{\delta \pJ(x)} + \frac{1}{2N^2}\frac{\delta^2}{\delta\pJ(x)^2} + \mathcal{O}(N^{-3})\;.
\end{align}
Consequently, the Master equation gives us the functional Fokker Planck equation for the density states when truncated at order $\mathcal{O}(N^{-2})$. Dealing first with lane switching, we have that
\begin{align}
    Q_{\lambda}(\vec{\phi},x) = &\mathcal{A}(\vec{\phi},x)+ \frac{1}{2N}\mathcal{B}(\vec{\phi},x)\;,
    \label{eq:functionallaneswitching}
\end{align}
where
\begin{align*}
    \mathcal{A}(\vec{\phi},x) = & \lambda\left(\frac{\delta}{\delta\pL(x)}-\frac{\delta}{\delta\pR(x)}\right)(\pR(x)-\pL(x))
\end{align*}
represents the drift in the density states associated with the lane switching and
\begin{align*}
    \mathcal{B}(\vec{\phi},x) = & \lambda\left(\frac{\delta}{\delta\pL(x)}-\frac{\delta}{\delta\pR(x)}\right)^2(\pR(x)+\pL(x))
\end{align*}
is a diffusive term.

Similar steps can be taken for the transport component of the Master equation. We can go further in this case by taking the limit $h\to0$ to represent continuous movement. In this limit it can be shown that (see Appendix) this term reduces to, perhaps unsurprisingly, an advective term in the Fokker Planck equation with no diffusive element. Specifically, we obtain
\begin{align}
    \iint Q^J_{h}(\vec{\phi},x,y)P(\vec{\phi})\text{d}x\text{d}y = \int\frac{\delta}{\delta\phi^J(x)}\partial_x(q^J)P(\vec{\phi})\d x
\end{align}
where $q^J=\phi^J(1-K\star\phi^J)$ is the flow rate for lane $J$. Hence, the full Fokker Planck equation is given by \eqref{eq:FullFokkerPlanck}.
\begin{widetext}
\begin{equation}
\begin{split}
    \partial_t P(\vec{\phi}) = & \int\frac{\delta}{\delta\phi^L(x)}\left\{\left[\lambda(\pR-\pL) - \partial_x(q^L)\right]P(\vec{\phi})\right\} + \frac{\delta}{\delta\phi^R(x)}\left\{\left[\lambda(\pL-\pR) - \partial_x(q^R)P(\vec{\phi})\right]P(\vec{\phi})\right\} \\ & \quad+ \frac{\lambda}{2N}\left(\frac{\delta}{\delta\pL(x)}-\frac{\delta}{\delta\pR(x)}\right)^2(\pR+\pL) P(\vec{\phi}) \d x \label{eq:FullFokkerPlanck}
\end{split}
\end{equation}
\end{widetext}

If the number of vehicles in the system, $N$, is sufficiently large we can neglect the second term in \eqref{eq:functionallaneswitching}. Doing so reduces the full Fokker Planck equation in \eqref{eq:FullFokkerPlanck} to a Liouville equation, for which
\begin{subequations}
\begin{align}
    \partial_t \phi^L & = \lambda(\phi^R-\phi^L) - \partial_xq^L \\
    \partial_t \phi^R & = \lambda(\phi^L-\phi^R) - \partial_xq^R
\end{align}
\end{subequations}
is a solution. Clearly the total density $\rho=\phi^L+\phi^R$ satisfies the continuity equation given by 
\begin{align}
    \partial_t \rho + \partial_x(q) = 0
\end{align}
where $q=q^L+q^R$. 

\subsection{Equispaced steady state}

Without lane changes, the vehicles will adjust their relative positions to each other until they become equally spaced. The speed of the vehicles in this state would be the assumed speed if lane changing had no influence. We study the average speed relative to this equidistant state flow which we can calculate exactly. In general, if there were $1\leq n\leq N$ vehicles in a lane, one would be lead to write 
\begin{align}
    \phi^*(x) = \frac{1}{N}\sum_{j=1}^n \delta\left(x - \frac{2\pi}{n}\right)\;.
    \label{eq:equiddensity}
\end{align}
We are specifically interested in the case where the lanes are equally populated ($n=N/2$) and we assume the total population is an even number for simplicity. Since the density in \eqref{eq:equiddensity} constitutes a periodic Dirac comb, the analysis can proceed in real or Fourier space. When we consider the fluctuations later a Fourier representation is analytically tractable and so we shall proceed similarly here. Defining $\phi_k$ as the $k^{th}$ Fourier mode of $\phi(x)$ and assuming the lanes are indeed equally split we find
\begin{align}
    \phi^*_k = \begin{cases}
        1/4\pi & k=mN/2\;,\quad m\in\mathbb{Z} \\
        0 & \text{otherwise}.
    \end{cases}
\end{align}
In other words, the Fourier representation is also a Dirac comb situated at every $N/2^{th}$ mode. The Fourier dual of the flow rate and hence the mean flow rate, $q_0$, are given by
\begin{subequations}
\begin{align}
    q_k & = \phi_k - 2\pi \sum_\ell K_\ell\phi_\ell\phi_{k-\ell} \label{eq:qmodes}\\
    q_0 & = \phi_0 - 2\pi \sum_\ell K_\ell |\phi_\ell|^2\;. \label{eq:q0}
\end{align}
\end{subequations}
The average velocity, $V$, is defined simply as the flow rate per individual or $V=q_0/\phi_0$. Thus, we have
\begin{align}
    V & = 1 - \frac{1}{2}\sum_{\ell=-\infty}^\infty K_{\ell N/2}\;.
\end{align}
By substituting in the Fourier modes of the coupling kernel
\begin{align}
    K_k = \frac{1}{2\pi}\frac{1+ik\alpha}{1+k^2\alpha^2}
    \label{eq:kernelfourier}
\end{align}
and recalling that $\alpha=4\pi m/N$, we can evaluate the sum exactly to give $V = 1 - \frac{\beta}{8\pi m}\coth\left(\frac{1}{2m}\right)$. However, this calculation has allowed self-interactions due to the nature of the convolution of the kernel with the density. To remove this undesirable effect and make it comparable to simulations we add on a correction term $K(0)/2N=\beta/8\pi m$ to account for the additional slow down that self-interactions would create:
\begin{align}
    V^* = 1 + \frac{\beta}{8\pi m}\left(1-\coth\left(\frac{1}{2m}\right)\right)\;.
\end{align}
Notice that this velocity would not have been the same had a homogeneous density, $\phi(x)=1/2\pi$, been assumed.

\section{Noise-induced slow down}

\begin{figure}
    \centering
    \includegraphics{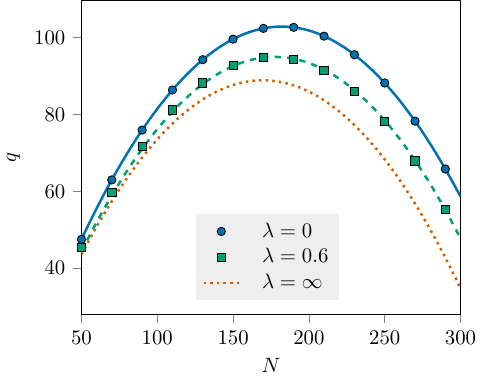}
    \caption{Fundamental diagram showing the decrease in flow when lane changes are allowed to occur and $\alpha$ is kept constant while the number of vehicles, $N$, change. Theoretical flow rate given by \eqref{eq:fundamentaldiagram} (with $q=V/\rho_0$) is shown when no lane changes can occur $(\lambda=0)$ and compared with the average flow when random lane changes, $\lambda=1.0$, reduce the flow. In all cases, we choose $\beta_0=0.04$ and $\alpha=\pi/25$.}
    \label{fig:fundamentaldiagram}
\end{figure}

The effect of a vehicle switching lanes causes a cascading effect on the vehicles behind which is most straightforward to capture in Fourier space. Instead of the functional Fokker Planck equation, we work with its Fourier series expanded counterpart. Thus we now have
\begin{align*}
    \pderiv{P}{t} =  -\sum_{J,k}\partial_{\pJ_k} \left(A^J_k P\right) + \frac{1}{2N}\sum_{I,J,k,\ell}\partial_{\phi^I_k} \partial_{\pJ_\ell}\left(B^{I,J}_{k,\ell} P\right)
\end{align*}
which represents the joint density of the Fourier modes $\phi^J_k$ where 
\begin{align*}
    A^L_k & = \lambda(\phi^R_k-\phi^L_k) - iku_k^L \\
    A^R_k & = \lambda(\phi^L_k-\phi^R_k) - iku_k^R
\end{align*}
and
\begin{align*}
    B^{I,J}_{k,\ell} & = (2\delta_{I,J}-1)\lambda(\phi^R_{k+l}+\phi^L_{k+l})\;.
\end{align*}
Following the approach of \cite{Rogers_2012,minors2018noise}, this Fourier Fokker Planck equation is equivalent to writing a set of SDEs for the Fourier modes of the density:
\begin{align}
\begin{split}
    \dot{\phi}^L_k & = \lambda(\phi^R_k-\phi^L_k) - ikq^L_k + \frac{1}{\sqrt{2\pi N}}\eta_k(t) \\
    \dot{\phi}^R_{k} & = \lambda(\phi^L_k-\phi^R_k) - ikq^R_{k} - \frac{1}{\sqrt{2\pi N}}\eta_{k}(t) 
    \label{eq:Forier2lanes}
\end{split}
\end{align}
where $\eta_k$ are complex Gaussian noises with correlation $\langle\overline{\eta_k(t)}\eta_l(t')\rangle = \lambda\delta(t-t')(\phi^L_{k+l}+\phi^R_{k+l}).$
\begin{figure}
    \centering
    \includegraphics{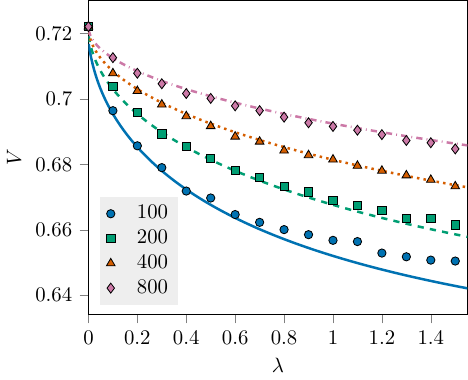}
    \caption{Average velocity, $V$, decreases as the rate of lane changing, $\lambda$ increases. Simulations performed for various system sizes $N$ are shown by points while the corresponding lines refer to the theoretical prediction in \eqref{eq:2lanevelocity}. Each simulation was carried out using the Euler-Maruyama method with $\beta=6$, $\alpha=4\pi/N$ until $T=500$ with step sizes of $10^{-3}$.}
    \label{fig:2lanes}
\end{figure}
We study the fluctuations around the equidistant density state by writing 
\begin{align}
    \xi^J_k = \sqrt{N}(\phi^J_k-\phi^*_k)
    \label{eq:linearflucs}
\end{align}
and assuming the fluctuations are symmetric $\xi_k=\xi^L_k=-\xi^R_k$. Applying this to \eqref{eq:q0}, the long time-averaged speed is
\begin{align}
    V = V^* - \frac{4\pi}{N}\sum_{\ell}K_\ell\langle|\xi_\ell|^2\rangle_\infty
    \label{eq:velocitylongtimeav}
\end{align}
where we use $\langle\dots\rangle_\infty$ to denote averaging according to the long-time stationary distribution of the density. Applying It\^o's formula to $\langle|\xi_k|^2\rangle$ using \eqref{eq:linearflucs} and \eqref{eq:Forier2lanes}, we obtain \eqref{eq:flucODEcorrelated}.
\begin{widetext}
\begin{align}
    \deriv{}{t}\langle|\xi_k|^2\rangle = & -\left(4\lambda+ikv\im(K_k)\right)\langle|\xi_k|^2\rangle - \frac{ikv}{2}\sum_{p\neq0}\left[\left(K_{np}+K_{-k-np}\right)\langle\xi_{-k-np}\xi_k\rangle-\left(K_{np}+K_{k-np}\right)\langle\xi_{k-np}\xi_{-k}\rangle\right] + \frac{\lambda\rho_0}{2\pi}
    \label{eq:flucODEcorrelated}
\end{align}
\end{widetext}
In practice, it can be seen that the correlation between modes, particularly for $N>100$, is negligible. By making the assumption that the modes are indeed uncorrelated, we obtain,
\begin{align}
    \deriv{}{t}\langle|\xi_k|^2\rangle = -\left[4\lambda+k\im(K_k)\right]\langle|\xi_k|^2\rangle + \frac{\lambda\rho_0}{2\pi}.
    \label{eq:flucsode}
\end{align}
Note, the result would have been identical had we expanded about a homogeneous density for which the Fourier modes are $\phi_k=\delta_{k,0}/4\pi$. From an initial state, the fluctuations evolve until they saturate at a value given by the steady state of \eqref{eq:flucsode}. This steady state is thus the long time average and is given by 
\begin{align}
    \langle|\xi_k|^2\rangle = \frac{\lambda\rho_0}{2\pi\left(4\lambda + k\im(K_k)\right)}\;.
\end{align}
We substitute this into \eqref{eq:velocitylongtimeav} and recall that $K_k=\bar{K}_{-k}$ so that the summation can now be expressed as 
\begin{align}
    V = V^* - \frac{1}{N}\left(\frac{K_0}{2}+4\lambda\sum_{k=1}^\infty \frac{\re(K_k)}{4\lambda + k\im(K_k)}\right)\;.
\end{align}
By substituting \eqref{eq:kernelfourier} in, the summation can be carried out exactly to obtain
\begin{align}
    V = V^* - \frac{\beta\kappa}{4N}\coth(\pi\kappa)
    \label{eq:2lanevelocity}
\end{align}
where $\kappa=\sqrt{8\pi\lambda/\alpha(8\pi\lambda \alpha+1)}$. For low vehicle numbers this decrease in average speed can be appreciable as shown in \Cref{fig:2lanes}. This effect can also been seen in the context of flow rates. If, instead of fixing the lengthscale $\alpha$ to the number of vehicles, we keep it constant and allow the number of vehicles to change we can see the effect of lane changing on the fundamental diagram (\Cref{fig:fundamentaldiagram}). To do this it is important to increase the coupling kernel strength linearly with $N$ since our model imposes a normalised density: $\beta=N\beta_0$. Leaving $\alpha$ free gives 
\begin{align}
    V = 1 + \frac{\beta_0}{2\alpha}\left(1-\coth\left(\frac{2\pi}{N\alpha}\right)\right) - \frac{\beta_0}{4}\coth(\pi\kappa')
    \label{eq:fundamentaldiagram}
\end{align}
where now $\kappa'=\sqrt{8\pi\lambda/\alpha(8\pi\lambda\alpha + \beta_0N)}$. As $\lambda$ increases, therefore, we still see this decrease in average flow. Interestingly, it is also evident that this effect becomes stronger as the density (number of vehicles) increases. As $\lambda\to\infty$, we find that $\kappa'\to\alpha^{-1}$ and the limiting value of the average speed is shown in \Cref{fig:fundamentaldiagram}.

\section{Three lanes}

It is reasonable to assume extending the analysis done thus far to three lanes would lead to a similar reduction in the average velocity across the lanes.
How the average velocity might vary between lanes is not immediately obvious since the middle lane has different behaviour to the side lanes.
Also, rather than having one source of noise we have two as the left and right lanes now behave independently. 
In order to maintain an equal distribution across the lanes, vehicles in the middle lane must change to either the left or right lanes at a rate $\lambda$ \textit{each}. 
The total flux at the stationary state in and out of the middle lane is thus twice that of the side lanes. 

By going through a similar process to the two-lane case can find an equivalent set of  SDEs for the Fourier modes of the three-lane system:
\begin{subequations}
\begin{align}
    \dot{\phi}^L_k & = \lambda(\phi^M_k-\phi^L_k) - ikq^L_k + \frac{1}{\sqrt{2\pi N}}\eta^L_k(t) \\
    \dot{\phi}^R_{k} & = \lambda(\phi^M_k-\phi^R_k) - ikq^R_{k} + \frac{1}{\sqrt{2\pi N}}\eta^R_{k}(t) \\
    \dot{\phi}^M_{k} & = \lambda(\phi^R_k+\phi^L_k-2\phi^M_k) - ikq^M_{k} + \frac{1}{\sqrt{2\pi N}}\eta^M_{k}(t)
\end{align}
\end{subequations}
where $\langle\overline{\eta^L_k(t)}\eta^L_l(t')\rangle = 2\lambda\delta(t-t')(\phi^M_{k+l}+\phi^L_{k+l})$ and similarly for $\eta^R_k$ but $\eta^M_k=-\eta^R_k-\eta^L_k$.

\begin{figure}
    \centering
    \includegraphics{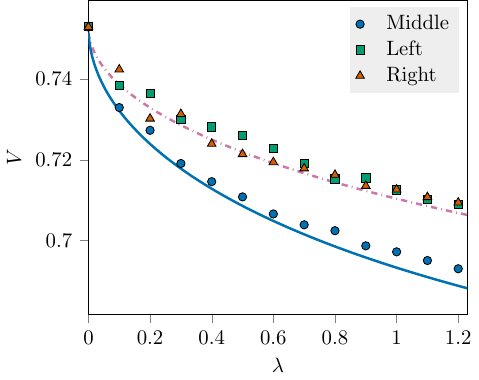}
    \caption{Average speed of each lane measured against rate of changing lanes. Time-averaged velocities from simulations of each lane are indicated by points while the corresponding lines refer to the theoretical prediction found from the expressions in \eqref{eq:2lanevelocity}. Each simulation was carried out using the Euler-Maruyama method with a system size of $N=600$, $\beta=8$, $\alpha=6\pi/N$ until $T=10^3$ with step sizes of $10^{-3}$.}
    \label{fig:3lanes}
\end{figure}

We assume the fluctuations in the middle lane can be characterised in terms of the fluctuations in the others:
\begin{align}
    \xi^{J}_k = \sqrt{N}(\phi^J_k - \phi^*_k)
\end{align}
and $\xi_k^M=-(\xi^L_k+\xi^R_k)$. Besides now being a two-component system, the setup is similar to the two-lane case and so we again assume the correlations between the Fourier modes are negligible. This is equivalent to assuming the homogeneous state $\phi^*_k=\delta_{k,0}/6\pi$ can be used about which the fluctuations exist to simplify the analysis. Consequently, we find that
\begin{subequations}
\begin{align}
    \deriv{}{t}\langle|\xi^L_k|^2\rangle = & -b_k\langle|\xi_k^L|^2\rangle - 2\lambda \re(\langle\xi^L_k\xi^R_{-k}\rangle) + \frac{2\lambda\rho_0}{6\pi} \\
    \deriv{}{t}\langle\xi^L_k\xi^R_{-k}\rangle = &-b_k\langle\xi^L_k\xi^R_{-k}\rangle -\lambda(\langle|\xi^L_k|^2\rangle+\langle|\xi^R_k|^2\rangle)
\end{align}
\end{subequations}
where $b_k=4\lambda - \frac{2}{3}k\im(K_k)$.
We only need to consider the real component of the cross fluctuations and by symmetry $\langle|\xi^L_k|^2\rangle=\langle|\xi^R_k|^2\rangle$. The stationary states for the expectation of these fluctuations are
\begin{subequations}
\begin{align}
    \langle|\xi_k^L|^2\rangle & = \frac{2\rho_0b_k\lambda}{6\pi(b_k^2-4\lambda^2)}\;, \\
    \re(\langle\xi_k^L\xi_{-k}^R\rangle) & = -\frac{4\rho_0\lambda^2}{6\pi(b_k^2-4\lambda^2)}\;.
\end{align}
\end{subequations}
The average velocity for the side lanes in terms of the fluctuations remained unchanged from before but the middle lane contains the cross correlation of the left and right lanes and is given by
\begin{align*}
    V^M & = V^* - \frac{2}{N}\sum_k K_k \left(\langle|\xi_k^L|^2\rangle + \langle|\xi_k^R|^2\rangle + 2\re(\langle\xi_k^L\xi_{-k}^R\rangle)\right)
\end{align*}
In this case, the summations do not reduce down to a simple expressions, but can still be evaluated by truncating the series. In \Cref{fig:3lanes} these results for the side and middle lanes can be seen with good agreement with particle simulations. As expected, the middle lane does, on average, have a lower velocity than the side lanes due to the extra flux in and out of the lane.

\section{Discussion}

In this article we have demonstrated the effect that spontaneous lane changing can have on free flow without bottlenecks. 
By taking a nonlocal model of traffic flow and allowing random lane changes to occur in the cases of two and three-lane traffic, we have quantified the overall effect on the average velocity of vehicles. 
While previous models have found a reduction in traffic flow when lane changes are considered, this often focuses around on-off ramps.
These models also assume well-behaved decision making of the individuals.
The approach taken here takes a different viewpoint, as the methodology allows for the analytic quantification of velocities for finite numbers of vehicles.
This enables the effect of individual random actions to be taken into account.

One of the main reasons often given for the reduction in flow due to increased lane changing intensity is the time taken for vehicles to change lanes, effectively making the vehicle in question occupy space in both lanes for a period of time.
This increased ``effective density'' thus contributes to a lower average velocity according to the fundamental diagram. 
Our model does not include this effect but instead suggests another perspective. The vehicle changing lanes causes a non-local effect on the density as trailing vehicles must adjust their velocity to avoid collision or crowding.
Of course it would be interesting to investigate the time taken to change lanes within our modelling framework; for example it may be possible to have a third species type (in the two lane model) which interacts with both lanes which all vehicles switch to briefly before completing a lane change.

Our model specification has intentionally been left minimal as we prefer to focus on the impact of unforced lane changes and the finite number of vehicles on roads.
Generalisations and additional factors such as informed lane changing decisions could be incorporated to give a more realistic depiction of driver behaviour.
However, while limiting the realism of the model, we highlight the potential impact such lane changes can have on the overall velocity.
It also exemplifies how careful consideration of the stochastic effects and system size needs to be taken.
Assuming a `hydrodynamic' limit of large numbers of vehicles, as is done to formulate continuum models, removes the additional velocity reduction found in this article.
The impact of finite population effects has been noted in myriad contexts, including phenotypic diversity \cite{Rogers_2012}, stochastic Turing patterns \cite{Biancalani,karig2018stochastic}, and the collective motion of general interacting particle systems \cite{worsfold23Density}.
Until now, quantifying the finite sized effects of traffic has been difficult even for single lane models.

\acknowledgments
JW supported by the EPSRC: EP/S022945/1 and the SAMBa Centre for Doctoral Training.


\section{Advective flow in the Master Equation}

Here we give a more detailed explanation of how the advective movement of vehicles along the road represented by $Q_h$ in the Master Equation results in a deterministic flow in the density evolution equation. Since this term is independent of neighbouring lanes, we drop the lane index to focus on the transport of one species $\phi(x)$ with instantaneous velocity $u(\phi,x)$. Under this simplification, the rate function is written as 
\begin{align*}
    Q_{h}(\vec{\phi},x,y) = (\Delta_x^{-}\Delta_y^{+}-1)d(x-y)\phi(x)
\end{align*}
where $d(x-y)=\frac{1}{h}\delta(x+u(\phi,x)h-y)$ is the jump rate density and $h$ is the step size. By expressing the step operators in terms of their functional Taylor series expansions up to second order we have 
\begin{align*}
    Q_h(\phi,x,y) = \mathcal{A}_h(\phi,x,y) + \frac{1}{2N}\mathcal{B}_h(\phi,x,y)
\end{align*}
with 
\begin{align*}
    \mathcal{A}_h(\phi,x,y) & = \left(\frac{\delta}{\delta\phi(y)}-\frac{\delta}{\delta\phi(x)}\right)d(x-y)\phi(x) \\
    \mathcal{B}_h(\phi,x,y) & = \left(\frac{\delta}{\delta\phi(y)}-\frac{\delta}{\delta\phi(x)}\right)^2d(x-y)\phi(x).
\end{align*}
In terms of their Fourier expansion, the first term becomes 
\begin{align*}
    \mathcal{A}_h(\phi,x,y) = \frac{1}{2\pi}\sum_{k,\ell}\partial_{\phi_k}\phi_\ell e^{i\ell x} \left(e^{-iky}-e^{-ikx}\right)d(x-y).
\end{align*}
Hence, the full integral is given by
\begin{align*}
    \iint\d x\d y \mathcal{A}(\phi)P(\phi) = & \lim_{h\to0}\frac{1}{2\pi h}\int\d x \sum_{k,\ell}\partial_{\phi_k}\phi_\ell e^{i\ell x} \\
    & \times\left(e^{-ik(x+uh)} - e^{-ikx}\right)P(\phi)
\end{align*}
Now, we expand the round brackets in orders of $h$:
\begin{align*}
    e^{-ik(x+uh)} - e^{-ikx} & = -ikuhe^{-ikx} + \mathcal{O}(h^2).
\end{align*}
As we take the limit $h\to0$, therefore, the integral above simplifies to 
\begin{align*}
    \iint\d x \d y & \mathcal{A}_h(\phi)P(\phi) \\ & = \frac{1}{2\pi}\int \d x \sum_{k,\ell} \partial_{\phi_k}\phi_\ell\left(-ikue^{i(p-n)x} \right)P(\phi)\\
    & = \sum_k \partial_{\phi_k}(-iku\phi_kP(\phi)).
\end{align*}
If the $N\to\infty$ limit is taken as before, this is the sole term which remains in the expanded Master Equation. Hence, this Liouville equation results in the following simple PDE for the density:
\begin{align*}
    \partial_t \phi + \partial_x(u\phi) = 0\;.
\end{align*}
What remains to be shown is that there are no extra, stochastic terms to be considered when a finite sized system is considered. To do this, we must show that $\lim_{h\to0}\mathcal{B}_h=0$. Explicitly, the Fourier expansion gives 
\begin{align*}
    \iint P\mathcal{B}_h & \d x\d y \\
    & = \frac{1}{4\pi^2}\iint \d x \d y\sum_{k,\ell,m}\partial_{\phi_k}\partial_{\phi_\ell}\phi_m e^{imx}d(x-y)\\ & \times\left[(e^{-i\ell y}-e^{-i\ell x})(e^{-iky}-e^{-ikx})\right]P(\phi)\;.
\end{align*}
Carrying out the integral in $y$ and focusing on the term in the square brackets, we find that
\begin{align*}
    & (e^{-i\ell(x-uh)} -e^{-i\ell x})(e^{-ik(x-uh)}-e^{-ikx}) \\ & = e^{-i(\ell+k)x}\left[1+e^{-i(k+\ell)uh}-e^{-i\ell uh}-e^{-ikuh}\right] \\
    & = iuhe^{-i(\ell+k)x}[n+m-(n+m)] \\
    & = 0
\end{align*}
as expected.

\end{document}